\documentclass[article]{aa}  
\usepackage{epsfig}
\usepackage{natbib}
\usepackage{txfonts}

\def\ms{\hbox{\,m\,s$^{-1}$}}         
\def\m2s2{\hbox{\,m$^{2}$\,s$^{-2}$}} 
\def\kms{\hbox{\,km\,s$^{-1}$}}       
\def\vsini{\hbox{$v$\,sin\,$i$}}      
\def\Msun{\hbox{$M_{\odot}$}}             
\def\Rsun{\hbox{$R_{\odot}$}}
\def\Mjup{\hbox{$M_{\rm Jup}$}}
\def\Rjup{\hbox{$R_{\rm Jup}$}}
\def\degr{\hbox{$^\circ$}}
\def\Bl{\hbox{$B_{\rm \ell}$}}
\def\Sca{\hbox{$S_{\rm Ca}$}}
\def\vrad{\hbox{$v_{\rm rad}$}}
\def\chisq{\mbox{$\chi^2$}}
\newcommand{\caii}{Ca$\;${\sc ii}}

\begin{document}

\title{Spectropolarimetric observations of the transiting planetary system of the K dwarf 
       HD~189733\thanks{Based on observations obtained at the Canada-France-Hawaii Telescope (CFHT) which is operated by the National Research Council of Canada, the Institut National des Sciences de l'Univers of the Centre National de la Recherche Scientifique of France,  and the University of Hawaii} }

\author{ C.~Moutou \inst{1}
\and J.-F.~Donati\inst{2}
\and R.~Savalle\inst{1}
\and G.~Hussain\inst{3}
\and E.~Alecian\inst{4,5}
\and F.~Bouchy\inst{6} 
\and C.~Catala\inst{4}
\and A.~Collier Cameron\inst{3} 
\and S.~Udry\inst{7} 
\and A.~Vidal-Madjar\inst{6}}

\offprints{\email{Claire.Moutou@oamp.fr}}

\institute{Laboratoire d'Astrophysique de Marseille, CNRS UMR 6110
               Traverse du Siphon, F-376 Marseille, France
\and
           Laboratoire d'Astrophysique Toulouse-Tarbes, Observatoire Midi Pyr\'en\'ees, 14 Av.\ E.~Belin, F-31400 Toulouse, France
\and
           School of Physics and Astronomy, University of St Andrews, North
               Haugh, St Andrews, Fife KY16 9SS, United Kingdom
\and
           LESIA, Observatoire de Paris Meudon, Place J. Janssen, 92195 Meudon cedex, France
\and
Dept. of Physics, Royal Military College of Canada, PO Box 17000, Stn Forces, Kingston, Canada K7K 7B4
\and
           Institut d'Astrophysique de Paris, UMR7095 CNRS, Universit\'e
               Pierre \& Marie Curie, 98bis Bd Arago, 75014 Paris, France              
\and
           Observatoire de Gen\`eve, 51 ch. des Maillettes, 1290
           Sauverny, Switzerland}

\date{Received / Accepted }

\abstract 
{With a Jupiter-mass planet orbiting at a distance of only 0.031~AU, the active K2 
dwarf HD~189733 is a potential candidate in which to study the magnetospheric 
interactions of a cool star with its recently-discovered close-orbiting giant planet. } 
{We decided to explore the strength and topology of the large-scale magnetosphere 
of HD~189733, as a future benchmark for quantitative studies for models of the 
star/planet magnetic interactions.  }  
{To this end, we used ESPaDOnS, the new generation spectropolarimeter at the 
Canada-France-Hawaii 3.6m 
telescope, to look for Zeeman circular polarisation signatures in the line profiles of 
HD~189733 in 2006 June and August. }
{Zeeman signatures in the line profiles of HD~189733 are clearly detected in all 
spectra, demonstrating that a field is indeed present at the surface of the star.  
The Zeeman signatures are not modulated with the planet's orbital period but apparently 
vary with the stellar rotation cycle.  
The reconstructed large-scale magnetic field, whose strength reaches a few tens of G, 
is significantly more complex than that of the Sun;  it involves in particular a 
significant toroidal component and contributions from magnetic multipoles of order up to 5.  
The \caii\ H \& K lines clearly feature core emission, whose intensity is apparently varying 
mostly with rotation phase.  Our data suggest that the photosphere and magnetic field of 
HD~189733 are sheared by a significant amount of differential rotation.  }
{Our initial study confirms that HD~189733 is an optimal target for investigating 
activity enhancements induced by closely orbiting planets.  More data 
are needed, densely covering both the orbital and rotation cycles, to investigate 
whether and how much the planet contributes to the overall activity level of 
HD~189733.  }

\keywords{stars: individual: HD189733 -- planetary systems --
     techniques: spectropolarimetry -- stars: magnetic field -- stars: activity }

\titlerunning{The magnetic field of Jupiter-mass planet hosting K dwarf HD~189733}

\authorrunning{C.~Moutou et al.}

\maketitle

\section{Introduction}

Planetary systems characterized by a giant planet at a few stellar radii from their
parent stars (known as hot Jupiters) currently make up $\sim$20\% of all known extrasolar planets.
Spectroscopic observations of stars hosting hot Jupiters recently revealed that
chromospheric activity indices (such as core emission in \caii\ H\&K and infrared triplet lines)
are sometimes modulated with the orbital period of the giant planet (rather than
with the rotation period of the star), suggesting that such giant planets may
significantly boost the activity level of the host star
\citep{cuntz00,shk03,shk05}.  
According to theoreticians \citep[e.g.,][]{cuntz00,ip04}, this 
modulation could result either from tidal 
effects (enhancing local dynamo action and activity at the surface of the star) or from 
magnetospheric interaction between the host star and its close-in planet (inducing reconnection 
events as the planet travels through the large magnetic loops anchored in the stellar surface).  
If due to reconnection, the resulting activity should depend mostly on the large-scale magnetic field 
of the host star, on the planet's magnetic field and on the orbital distance with respect to the 
Alfven radius of the host star;  if due to tidal effects, the excess activity may indicate the
presence of a ``surface wave'' of small-scale magnetic features comoving with the planet orbital motion 
and resulting from local turbulence and dynamo action enhancements induced by the planet pass-by.  
Photometric observations by the MOST space telescope of several hot-Jupiter systems suggest 
that stellar surface activity, in the form of cool spots comoving with the planet orbital motion, 
could indeed be caused by the close-in giant planet \citep{walker};  if confirmed, 
it would provide additional evidence for magnetic interactions between giant planets 
and their host stars.
Magnetic fields were also recently invoked as a possible formation mechanism for giant 
planets in close orbits around Sun-like stars \citep{kurucz07}.

Very recent observations of $\tau$~Boo with the ESPaDOnS spectropolarimeter 
\citep{donati97, donati07}, mounted at the Cassegrain focus of the 
3.6m Canada-France-Hawaii Telescope (CFHT), revealed that surface magnetic fields are indeed 
present at the surface of planet-hosting stars and can be reliably detected through the 
Zeeman signatures they generate in spectral line profiles \citep{catala07}.  Despite a sparse 
temporal sampling of the rotational cycle, their data demonstrated the feasibility of measuring 
and mapping large-scale magnetic fields at the surfaces of such stars, opening radically new and 
extremely promising perspectives for observational studies of the magnetic interactions
between active stars and their close-in giant planets.  We therefore decided to initiate a new 
program aimed at detecting and modeling the magnetic topologies of stars hosting close-in giant 
planets, in order to provide theoretical models of planet-star magnetic interaction with quantitative 
constraints.  With its short orbital period, the transiting planetary system HD~189733, discovered 
at Observatoire de Haute Provence through both photometric and spectroscopic signatures 
\citep{bouchy05}, is an ideal candidate for our program.

Most parameters of HD~189733 are now tightly constrained, thanks to successive very detailed 
spectroscopic and photometric monitoring programs \citep{bouchy05,bakos06,winn07a,winn07b}.  
The planet's mass and radius are equal to $1.13\pm0.03$~\Mjup\ and 
$1.156\pm0.046$~\Rjup\ respectively, while its orbital period is 
$2.2185733\pm0.0000019$~d.  
The system geometry is also well known, the inclination $i$ of the orbital axis with respect to the 
line of sight being $85.76\pm0.29\degr$.  With a mass, effective temperature and radius 
respectively equal to $0.82\pm0.03$~\Msun, $5050\pm50$~K and $0.753\pm0.023$~\Rsun, the K2V parent 
star is slightly metal poor and active, with a line-of-sight projected equatorial rotation velocity 
$\vsini\ =3.0\pm0.2$~\kms. 
In addition to the photometric eclipses induced by the transiting exoplanet, HD~189733 exhibits
intrinsic photometric variability;  this is likely the result of its activity and the 
dark spots on its surface, modulating the amount of light received from the star as rotation 
carries spots in and out of the observer's view \citep{hebrard06, winn07b, matthews07, croll07}.  
From Hipparcos observations collected over 3~yr, \citet{hebrard06} confirm that intrinsic 
variability is indeed the main source of out-of-eclipse photometric fluctuations on HD~189733, 
and that these fluctuations are modulated with an average rotation period of 11.8~d.  
From 93 nights of ground-based photometry secured from 2005 October to 2006 July, \citet{winn07b} 
show that the light curve of HD~189733 is evolving significantly on a timescale of 20~d.  Using their 
last 30~d of observations (showing the highest amplitude modulation), they derive a rotation 
period of $13.4\pm0.4$~d, significantly longer than that of \citet{hebrard06}.  Nearly-continuous 
space-based photometry of HD~189733 was obtained over 21~d in July and August 2006 with MOST, 
only 50~d after the latest observations of \citet{winn07b};  from these data, the rotation 
period was found to be 11.8~d to within better than 0.1~d \citep{matthews07, croll07}.  
We suspect 
that the discrepancy between these different estimates is not due to measurement errors, but rather 
to temporal variations in the spot pattern coupled to a significant amount of differential 
rotation over stellar latitudes \citep[e.g.,][]{hall91, catala07}.  We return to this issue
later in the 
paper;  in the meantime, we adopt an average rotation period of 11.8~d for HD~189733.  
Table~\ref{tab:par} summarises all relevant system properties that are used in this paper and 
corresponding references.  

This paper presents the first spectropolarimetric study of HD~189733.  Our aim is 
to detect and monitor the Zeeman signatures in the spectrum of the host star to map 
its large-scale magnetic topology and study the magnetic interaction between the star and 
its close-by giant planet.  In Section~\ref{sec:obs} we describe the observations, while 
in Sect.~\ref{sec:mod} we detail the modeling of the spectropolarimetric data.  
We discuss the activity level of the star and its possible variation 
with orbital and rotational phase in Section~\ref{sec:act}.  In Sect.~\ref{sec:atm} we examine 
spectra taken during transit for potential spectroscopic signatures of the planet's atmosphere, and in
Sect.~\ref{sec:ref} we derive upper limits on the detection of starlight reflected by the planet.  
Finally, in Sect.~\ref{sec:dis} we summarise the results , discuss their implications and suggest 
directions to extend our study in the future.

\begin{table}
\begin{center}
\caption[]{Star and planet parameters (and associated error bars) 
for the transiting system HD~189733. Corresponding 
references are: $^a$ \citet{bouchy05}, $^b$ \citealt{bakos06}, 
$^c$ \citealt{winn07a}, $^d$ \citealt{croll07}. }  
\begin{tabular}{ll}
\hline
Planet mass (\Mjup) $^c$ & 1.13 (0.03)\\
Planet radius (\Rjup) $^b$& 1.154 (0.032)\\
Orbital period (d) $^c$ &  2.2185733 (0.0000019)\\
Semi-major axis (AU) $^c$& 0.031 (0.001)\\
Stellar mass (\Msun) $^a$& 0.82 (0.03) \\
Stellar radius (\Rsun) $^c$& 0.753 (0.02) \\
Stellar effective temperature (K) $^a$& 5050 (50) \\
Stellar \vsini\ (\kms) $^c$& 2.97 (0.22) \\
Stellar rotation period (d) $^d$& 11.73 (0.07)\\
Orbit inclination (\degr) $^b$& 85.76 (0.29)\\
\hline
\end{tabular}
\end{center}
\label{tab:par}
\end{table}

\section{Observations}
\label{sec:obs}

Spectropolarimetric data of HD~189733 were collected with ESPaDOnS and CFHT, during 2 
separate runs in 2006 June (10--13) and August (05--12).  With ESPaDOnS, the complete 
optical spectrum (from 370 to 1000~nm) is collected in a single exposure at a spectral 
resolution of about 65,000;  from sequence of 4 subexposures taken in different configurations 
of the polarimeter, one can obtain both the intensity and polarisation spectrum, in 
either circular (Stokes $V$) or linear (Stokes $Q$ or $U$) polarisation state 
\citep{donati97,donati07}.  
Circular polarisation spectra give access to the Stokes $V$ Zeeman signatures and thus 
to the longitudinal field component averaged over the visible stellar hemisphere;  
they allow us to retrieve the parent large-scale magnetic field at the surface of the 
star  (see Sec.~\ref{sec:mod}).  
Linear (Stokes $Q$ and $U$) polarisation can potentially inform us of the amount of 
stellar light reflected off the dusty planetary atmosphere;  this signal (whose typical 
size is of order a few tens of ppm with respect to the stellar flux and depends on the 
planet albedo) is expected to be modulated by the planet orbital phase (see 
Sec.~\ref{sec:ref}) and to reach its maximum size near the planet orbital quadrature. 
Eight Stokes $V$ sequences at 3 main epochs, 
and 7 Stokes $V$ sequences at 6 main epochs were collected during the first and second run 
respectively, the 4 remaining sequences corresponding to Stokes $Q$ and $U$ sequences. 
Signal to noise ratios (S/N) range from 600 to 1100 per 2.6~\kms\ velocity bin, with 
typical values between 900 and 1000.  
The complete journal of observations is given 
in Table~\ref{tab:log}.  

Orbital and rotational phases (listed in Table~\ref{tab:log}) were computed
respectively using the 
ephemerides (see \citealt{winn07b,croll07}):   

\begin{equation}
T_0 = \mbox{HJD~}2,453,988.80336 + 2.2185733 E_{\rm orb}
\label{eq:orb}
\end{equation}
\begin{equation}
T_0 = \mbox{HJD~}2,453,988.80336 + 11.8 E_{\rm rot}
\label{eq:rot}
\end{equation}

A full planet transit was observed on June 13, with sequences 10 and 11 corresponding to 
transit (ie orbital phase equal to 0.0$\pm0.017$), while sequences 9 and 12 correspond to 
phase ranges immediately preceding and following transit.

The raw data were reduced using Libre-ESpRIT, the dedicated and fully automatic 
spectropolarimetric reduction package 
available to ESPaDOnS users at CFHT \citep{donati97,donati07}.  
From collected calibration exposures and stellar frames, Libre-ESpRIT automatically extracts 
wavelength-calibrated intensity and polarisation spectra with associated error bars at each 
wavelength pixel, shifted to the heliocentric rest frame and normalised to a unit continuum.  
In particular, all spectra are automatically corrected for spectral shifts resulting from 
instrumental effects (e.g. mechanical flexure, temperature or pressure variations) using 
the numerous telluric lines present in each spectra as a stable wavelength reference.  
Through cross-correlation, we estimate the relative shift of telluric lines with respect to 
the observer's frame;  we assume that this shift is entirely attributable to instrumental 
effects and simply remove it by translating the spectra and recenter the telluric lines in 
the observer's rest frame.  This is of course only an approximation;  telluric lines are 
actually not a perfectly stable reference in the observer's rest frame, eg due to the effect of strong 
winds in the upper atmosphere (where telluric lines form) that can potentially shift these 
lines by as much as a few tens of \ms.  In practice, we find that our method yields a 
radial velocity differential rms precision of about 15~\ms\ within a single run, with 
systematic shifts of up to 30~\ms\ from run to run \citep{donati07}.  This method is less 
accurate than the classical technique which consists in recording the spectrum of a reference 
ThAr lamp simultaneously (and interleaved) with that of the main star, yielding typical 
differential radial velocity accuracies of a few \ms;  our method is nonetheless accurate enough to 
detect the planet-induced radial velocity variations of HD~189733 (see below) and to ensure 
a reliable rotational modulation analysis of the detected Zeeman signatures (see Sec.~\ref{sec:mod}).  

\begin{figure}
\label{fig:lsd}
\begin{center}
\epsfig{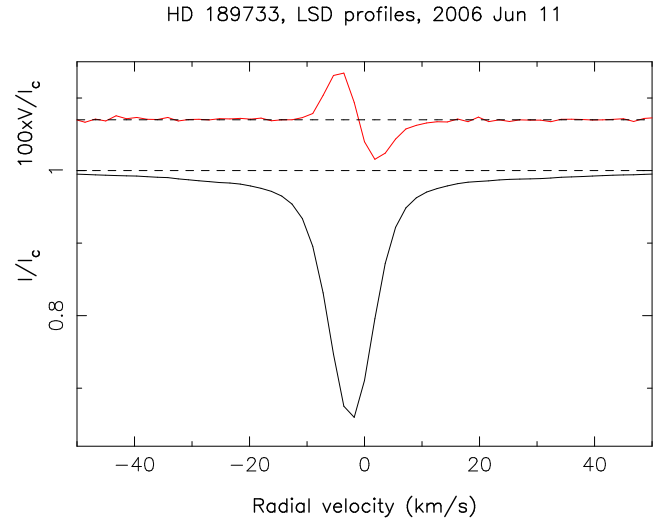}
\caption{LSD unpolarised (bottom) and circularly polarised (top) profiles of HD~189733 
on 2006 June 11.  All LSD profiles obtained on this night (corresponding to observation 
sequences 3 to 5, see Table~\ref{tab:log}) were averaged together, further decreasing 
the relative noise level (in the polarisation profile) down to 19~ppm.  In this particular 
example, the full amplitude of the Stokes $V$ signal is 0.12\%, ie 60 times greater than the rms 
noise level.  A Zeeman signature is obviously detected, corresponding to a longitudinal 
field value of about 8~G.  Note that the Stokes $V$ profile was expanded by 100 and 
shifted upwards for graphics purposes.  } 
\end{center}
\end{figure}

Least-Squares Deconvolution \citep[LSD, ][]{donati97} was applied to all spectra, using a line 
pattern derived from a Kurucz model atmosphere with solar abundances, and effective temperature 
and logarithmic gravity set to 5000~K and 4.5 respectively.  This line pattern includes most 
moderate to strong lines present in the optical domain (those exhibiting central depths 
larger than 40\% of the local continuum, before any macroturbulent or rotational broadening,  
about 7,000 throughout the whole spectral range) but excludes the very 
strongest, broadest features 
such as the Balmer lines, whose Zeeman signature is strongly smeared out
compared to those of narrow lines.  The output of LSD for each sequence of 
polarisation spectra is a pair of average 
line profiles: either 
intensity (Stokes $I$) and circular (Stokes $V$) polarisation profiles,
or linear  (Stokes $Q$ and $U$) polarisation profiles, depending on the 
type of observation sequence. We will refer to these as LSD profiles throughout the paper.   The 
typical multiplex gain in S/N for polarisation profiles is 30, implying noise levels in LSD 
polarisation profiles as low as 30 to 40~ppm (ie 3 to $4\times10^{-5}$ in units of the 
unpolarised continuum) depending essentially on the input spectrum quality.  An example 
of such profiles is shown in Fig.~\ref{fig:lsd}.  

Stokes $V$ Zeeman signatures are detected for all circular polarisation observations (15 
profiles altogether).  From the first moment of the Stokes $V$ LSD profiles, we derived 
estimates of the line-of-sight component of the magnetic vector averaged over the visible 
stellar hemisphere, ie the so-called longitudinal field \Bl\ \citep{donati97}.  The values 
obtained are listed in Table~\ref{tab:log} and typically range from $-8$ to $+8$~G.  Note 
that longitudinal field values only represent a small fraction of the information stored in 
Stokes $V$ profiles. For example, while the Stokes $V$ signature corresponding
to sequence 16 is clearly detected, featuring a signal with a full amplitude
of 0.06\% (ie 20 times larger than the corresponding error bar), the
corresponding longitudinal field value is consistent with  0~G within the noise level and is 
thus unable to diagnose properly the field detection we obtained.  Nonetheless, longitudinal-field 
values are convenient summary statistics in 
several respects, and in particular can provide useful estimates of the average 
magnetic flux over the stellar surface when the field polarity is roughly uniform over 
the visible stellar hemisphere.  In particular, these values indicate straightforwardly 
(and not unexpectedly) that \Bl\ correlates very poorly with orbital phase. Sequences 3 
and 12, for example, correspond to very similar orbital configurations, but yield widely discrepant \Bl\ 
values. On the other hand, \Bl\ varies rather smoothly with stellar rotation phase, changing from positive to 
negative values throughout the first observing run.  A more complete modeling of the 
magnetic topology from our collection of LSD Stokes $V$ profiles is presented in 
Sect.~\ref{sec:mod}.  

\begin{figure}
\begin{center}
\epsfig{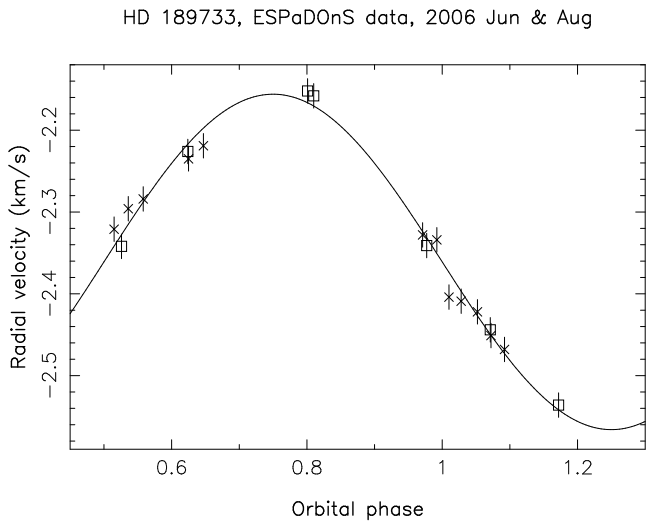}
\caption{Radial velocities of HD~189733 derived from ESPaDOnS spectra collected 
in 2006 June (crosses) and August (squares) as a function of orbital phase (using 
the ephemeris of eqn.~\ref{eq:orb}).  The radial velocity model of \citet{bouchy05} 
is also plotted (full line).  Note that the measurements were shifted by $-0.13$~\kms\  
and $-0.23$~\kms\ for the June and August data respectively, to compensate for a 
global shift with respect to the model predictions (see text).  Individual $1-\sigma$ 
error bars of 15~\ms\ are also plotted.  Note that the deviating point near phase 1.0 
is compatible with the Rossiter-Mc Laughlin effect \citep{bouchy05,winn07b}. } 
\end{center}
\label{fig:vrad}
\end{figure}

From Gaussian fits to each LSD Stokes $I$ profile, we derive radial velocities of 
HD~189733 throughout our 2 runs.  
The values we obtain (listed in Table~\ref{tab:log}) 
are in good agreement with expectations (using the orbital solution of \citealt{bouchy05} 
and the ephemeris of \citealt{winn07b} given in eqn.~\ref{eq:orb}) once the data are globally 
shifted by $-0.13$~\kms\ and $-0.23$~\kms\ for the June and August sets respectively 
(see plot on Fig.~\ref{fig:vrad}).  Residuals to the model are equal to 15 and 10~\ms\ rms 
for the June and August run respectively, confirming that this is indeed the typical 
radial velocity differential accuracy that ESPaDOnS can reach.
While the global shift of about 0.2~\kms\ 
between the ESPaDOnS and the ELODIE data \citep{bouchy05} may be attributable to 
differences in the radial velocity calibration procedures between the 2 instruments, 
the 100~\ms\ relative shift between our June and August data sets is significantly larger 
than the upper limit corresponding to the long-term stability of ESPaDOnS (about 30~\ms).  
It suggests that HD~189733 could host more than one giant planet and feature at least another 
one at a typical distance of 0.5~AU.  This will be investigated through future 
long-term spectroscopic monitoring.  

Finally, we perform a bisector analysis from normalised LSD Stokes $I$ profiles and measure 
the bisector span, from average velocities corresponding to $10-40$~\% and $55-85$~\% of the full 
LSD profile depth. The error bar on the bisector span is generally of the order of the error bar 
on the radial velocity estimate, about 15 m/s in the present case.

\begin{table*}
\begin{center}
\caption{Journal of spectropolarimetric observations.  The successive columns list the sequence number 
(col.~1),  the UT date (col.~2), 
the UT Heliocentric Julian date (col.~3), the orbital phase (using ephemeris of eqn.~\ref{eq:orb}, 
col.~4), the rotational phase 
(using ephemeris of eqn.~\ref{eq:rot}, col.~5), the measured polarisation state (col.~6), the total exposure time (col.~7), 
the peak signal to noise ratio in the spectrum (per 2.6~\kms\ velocity bin, col.~8), the longitudinal magnetic field \Bl\ 
whenever Stokes $V$ is measured (with corresponding 1$\sigma$ error bars, col.~9), the equivalent width \Sca\ of the residual 
emission core in the \caii\ H\&K line (col.~10, see Sect.~\ref{sec:act}) and the radial velocity \vrad\ (col.~11).}
\begin{tabular}{lllllllllll}
\hline
seq.\ & date (UT)  & HJD        & $\phi_{\rm orb}$ & $\phi_{\rm rot}$ & Stokes & exp.~time & peak S/N            & \Bl & \Sca & \vrad   \\
(\#)         &         &(2,453,000+)&                  &                  &        &  (sec)        & (per pixel) & (G) & (\kms) & (\kms) \\       
\hline
 1 & 2006 Jun 10 &897.01115 & 0.625 & 0.221 & $Q$ & 4$\times$960  & 980  &             & $-0.56$ & $-2.105$\\
 2 & 2006 Jun 10 &897.05805 & 0.647 & 0.225 & $U$ & 4$\times$960  & 1040 &             & $-0.82$ & $-2.089$\\
 3 & 2006 Jun 11 &897.95814 & 0.052 & 0.301 & $V$ & 4$\times$900  & 1080 & $+7.2\pm0.6$& $-0.11$ & $-2.292$\\
 4 & 2006 Jun 11 &898.00269 & 0.072 & 0.305 & $V$ & 4$\times$900  & 1000 & $+7.7\pm0.7$& $-0.53$ & $-2.321$\\
 5 & 2006 Jun 11 &898.04669 & 0.092 & 0.309 & $V$ & 4$\times$900  & 930  & $+7.9\pm0.8$& $-0.16$ & $-2.338$\\
 6 & 2006 Jun 12 &898.98463 & 0.515 & 0.388 & $V$ & 4$\times$900  & 830  & $+2.4\pm0.8$& $+0.20$ & $-2.191$\\
 7 & 2006 Jun 12 &899.03159 & 0.536 & 0.392 & $Q$ & 4$\times$1000 & 1040 &             & $+0.27$ & $-2.166$ \\
 8 & 2006 Jun 12 &899.08036 & 0.558 & 0.396 & $U$ & 4$\times$900  & 1140 &             & $+0.11$ & $-2.154$\\
 9 & 2006 Jun 13 &899.99747 & 0.971 & 0.474 & $V$ & 4$\times$730  & 890  & $-8.5\pm0.8$& $+0.43$ & $-2.198$\\
10 & 2006 Jun 13 &900.04204 & 0.992 & 0.478 & $V$ & 4$\times$846  & 1000 & $-6.3\pm0.7$& $+0.55$ & $-2.204$\\
11 & 2006 Jun 13 &900.08354 & 0.010 & 0.481 & $V$ & 4$\times$846  & 960  & $-8.3\pm0.7$& $+0.73$ & $-2.274$\\
12 & 2006 Jun 13 &900.12387 & 0.028 & 0.485 & $V$ & 4$\times$600  & 790  & $-6.9\pm0.9$& $+0.36$ & $-2.279$\\
\hline
13 & 2006 Aug 05 &952.86578 & 0.801 & 0.954 & $V$ & 4$\times$350  &  620 & $+2.1\pm1.1$& $-0.57$ & $-1.922$\\         
14 & 2006 Aug 05 &952.88433 & 0.810 & 0.956 & $V$ & 4$\times$275  &  560 & $+0.9\pm1.3$& $-0.99$ & $-1.928$\\         
15 & 2006 Aug 08 &955.90633 & 0.172 & 0.212 & $V$ & 4$\times$715  &  900 & $-4.1\pm0.8$& $-0.20$ & $-2.306$\\      
16 & 2006 Aug 09 &956.90905 & 0.624 & 0.297 & $V$ & 4$\times$1050 &  1120& $+0.2\pm0.6$& $-0.69$ & $-1.996$\\     
17 & 2006 Aug 10 &957.90042 & 0.071 & 0.381 & $V$ & 4$\times$900  & 1050 & $+2.1\pm0.7$& $+0.32$ & $-2.214$\\ 
18 & 2006 Aug 11 &958.91094 & 0.526 & 0.467 & $V$ & 4$\times$880  & 1030 & $+3.2\pm0.7$& $+0.75$ & $-2.112$\\ 
19 & 2006 Aug 12 &959.91103 & 0.977 & 0.551 & $V$ & 4$\times$880  &  980 & $+2.8\pm0.7$& $+0.78$ & $-2.111$\\ 
\hline
\end{tabular}
\end{center}
\label{tab:log}
\end{table*}

\section{Modeling the magnetic topology of HD~189733}
\label{sec:mod}

To model the magnetic topology of HD~189733, we use our magnetic-imaging code 
\citep{brown91, donati97b} in its 
most recent implementation \citep{donati06}.  While still 
based on maximum-entropy image
reconstruction, this latest version reconstructs the field topology as
a spherical-harmonic decomposition, rather than as a series of
independent magnetic-image pixels.  One obvious advantage of
this method is that we can impose {\em a priori} constraints on the
field topology -- e.g., that the field is purely poloidal, or purely
toroidal, or more generally, a combination of both.  Another important advantage of
this formalism is that both simple and complex magnetic topologies can
easily be reconstructed \citep{donati06}, whereas the original method
failed at reconstructing simple magnetic geometries \citep[such as
dipoles;][]{brown91}.  Details about how we effected this approach are 
given in \citet{donati06}.  In this section, we assume that the Stokes $V$ profiles 
are modulated with the rotation period of 11.8~d;  we further assume that the 
rotation spin axis of the star coincides with the orbital axis, inclined at 85.7\degr\ 
with respect to the line of sight\footnote{
While \citet{winn07b} find that the 
spin and orbit axis are most likely aligned in HD~189733, \citet{croll07} conclude 
in favour of a moderate spin-orbit misalignment from their very accurate photometric 
data set.  However, the latter mention that their result only holds if no differential 
rotation is present at the surface of the star, an assumption whose validity is fairly 
uncertain (see end of Sec.~\ref{sec:mod}).  We thus assume here that the spin and orbit are 
aligned;  a 20\degr\ change in the assumed inclination has however very limited 
impact on the derived magnetic topology.  

}.  

To describe the LSD Stokes $V$ profiles of HD~189733, we use the same simple model 
of \citet{donati97} and \citet{donati06}.  
In this model, the local unpolarised line profile is described 
with a Gaussian, whose full-width-at-half-maximum is set to 7~\kms\ and whose 
equivalent width matches that of the observed LSD Stokes $I$ profiles (3.2~\kms);  
integrating over the stellar surface and taking $\vsini=3.0$~\kms\  
\citep{winn07b} enables us to reproduce the observed Stokes $I$ profiles fairly well, once the model 
profiles are broadened with an instrumental broadening of 4.6~\kms.  We then assume 
that the local Stokes $V$ profile is proportional to the first derivative of the 
local Stokes $I$ profile (weak field approximation), the scaling factor 
depending in particular on the local longitudinal field component and the Land\'e factor of the 
average line (set here to 1.25).  For more details, the reader should refer to 
\citet{donati97}.  

Given the fairly low $\vsini$ of HD~189733, it is obvious that the spatial resolution 
we can achieve at the surface of the star is rather limited.  From the width of the 
local line profile (about 7~\kms), we derive that there are about 3 resolution elements 
across the stellar equator, implying that using spherical harmonics expansions with $\ell\leq5$ 
is sufficient for our needs.  
It also implies that most small-scale magnetic features potentially present at the surface of 
HD~189733, e.g., in the form of small bipolar spot groups like those peppering the surface of the 
Sun, will remain totally undetected in spectropolarimetric data sets;  this reflects the 
fact that close-by opposite magnetic polarities mutually cancel their respective effect in the 
Zeeman polarised signatures, that are sensitive to the vector properies of the field.  

\begin{figure*}
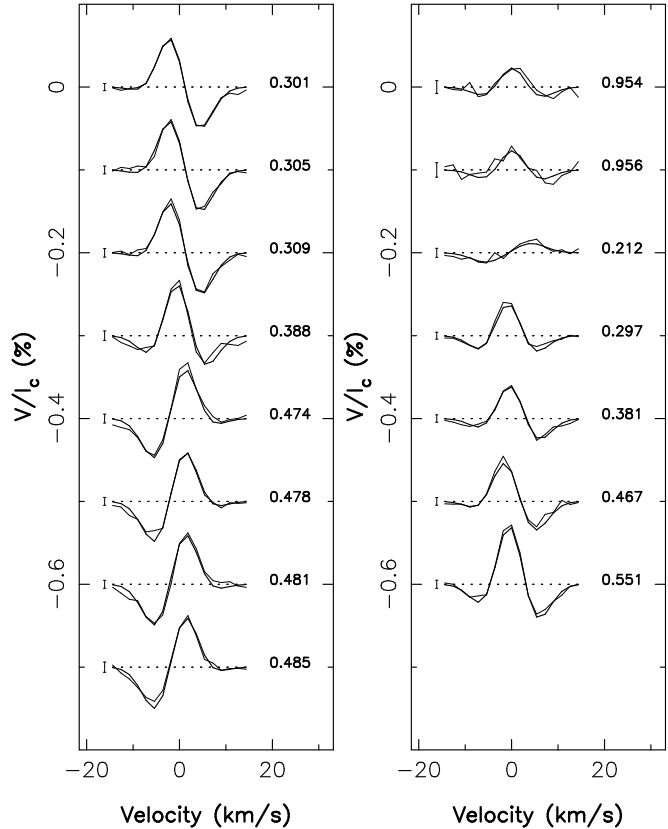

\begin{center}
\epsfig{file=moutou_fig3a.ps,width=11cm,angle=270}
\epsfig{file=moutou_fig3b.ps,width=11cm,angle=270}
\caption[]{LSD Stokes $V$ profiles of HD~189733 (thin line) and corresponding maximum-entropy fit 
(thick line) for our 2006 June (left panel) and August (right panel) data sets.  The rotational 
phases of observations (as derived from the ephemeris of eqn.~\ref{eq:rot}) 
are noted to the right of each profile.  A 3-$\sigma$ error bar is also depicted 
to the left of each profile.}
\end{center}
\label{fig:fit}
\end{figure*}

\begin{figure*}
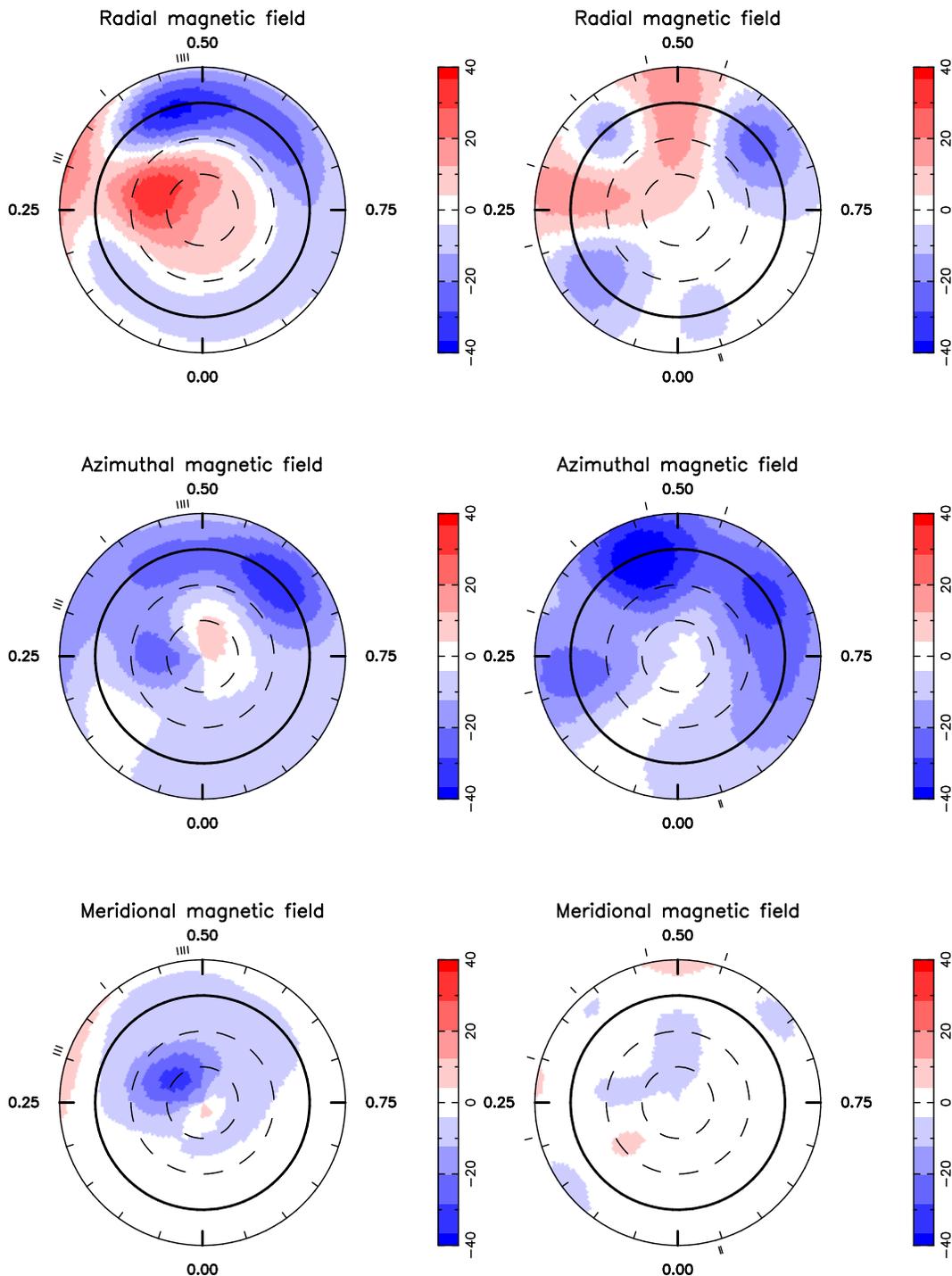

\begin{center}
\epsfig{file=moutou_fig4a.ps,width=7cm}
\epsfig{file=moutou_fig4b.ps,width=7cm}
\caption[]{Maximum-entropy reconstructions of the magnetic topology of
HD~189733 as derived from our 2006 June (left panel) and August (right panel) 
data sets, assuming that the global field can be expressed as the sum of a
poloidal field and a toroidal field.  The three components of the field
are displayed from top to bottom (flux values labelled in G).
The star is shown in flattened polar projection down to latitudes
of $-30\degr$, with the equator depicted as a bold circle and
parallels as dashed circles.  Radial ticks around each plot indicate
phases of observations (as derived from the ephemeris of eqn. 2). } 
\end{center}
\label{fig:map}
\end{figure*}

The maximum-entropy fits we obtain in this context for each of 
our 2 data 
sets are shown on Fig.~\ref{fig:fit}, while the corresponding maps (assuming a magnetic field 
featuring both poloidal and toroidal components) are shown on Fig.~\ref{fig:map}.  
Both data sets are fitted down to the noise level and 
correspond to a reduced \chisq\ of order unity.  Note that in both cases, the data set is only 
fragmentary, covering no more than 20\% and 50\% of the rotation cycle for the 
June and August sets respectively.  
It explains in particular why only little magnetic flux is reconstructed at the stellar surface 
in regions not (or only marginally) constrained by observations.  This is due to the fact that the 
imaging code is tailored to produce the simplest surface magnetic topology compatible with the 
data;  it therefore spontaneously biases the solution towards topologies featuring little magnetic 
flux on poorly observed regions of the stellar surface, provided that (i) this is compatible with 
observations at other rotational phases and that (ii) the assumed field topology is complex enough 
to allow a mixture of non-magnetic and magnetic regions (i.e., the spherical harmonics expansion 
describing the field topology includes orders significantly larger than 1).   
For this reason, we emphasise that the modeling we carry out in this paper is only preliminary 
and deserves further confirmation and assessment through more complete data sets.  However,
a number of reliable conclusions can be derived already from the present data.  

The first point we can address concerns the topology of the field.  By trying to fit the data 
with different field configurations (eg a poloidal field only, or a general poloidal plus toroidal 
field combination), one can check which topology is more likely to be present at the surface of 
the star. In the present case, we find that the field of HD~189733 most probably features both
a poloidal and a toroidal component at the surface of the star. Although both options give equivalent 
likelihoods in terms of the quality of fit to the data, the inclusion of a toroidal field component 
produces a magnetic configuration with significantly lower contrast (hence greater entropy and higher 
prior probability), yielding a higher posterior probability.  This is particularly obvious for the 
August data set, for which the 
poloidal plus toroidal field configuration 
that fits the data at the noise level is almost twice as weak on average as that obtained when 
assuming that the star hosts a purely poloidal field (not shown here).  This conclusion can be inferred
straightforwardly from the 
corresponding Stokes $V$ data set (see Fig.~\ref{fig:fit}).  At this epoch, the mean Stokes $V$ 
profile averaged over the whole series, as well as more than half of the individual profiles 
(eg that at phase 0.620), are clearly more-or-less symmetric about the line center. This is the 
characteristic signature of an azimuthal field ring encircling most of the star 
\citep[eg, ][]{petit05}.  
The same conclusion may be deduced from the June data set, although the evidence is 
weaker than for the other (more complete) data set.  In particular, the portions of the azimuthal 
field ring reconstructed at both epochs feature the 
same (ie clockwise or negative) polarity (see 
Fig.~\ref{fig:map}), strengthening the evidence that this toroidal 
field ring is indeed a real component 
of the magnetic topology of HD~189733.  

We can also be fairly confident that the magnetic field at the surface of the star is significantly 
more complex than that of the Sun.  
Magnetic field models with a spherical harmonic expansions restricted 
to $\ell\leq2$ produce significantly worse fits to the August data set, with reduced \chisq\ values 
always larger than 2.5.   

The modulation of the 
Zeeman signature on the August run looks very different from the regular shift from one 
main polarity to the other in about half 
a rotation period (as observed from most chemically peculiar stars, \citealt{wade00})
that  a simple dipolar-like 
topology would produce. Similarly, 
we observe in June a complete transition from one polarity to another in no more than 15\% of the 
rotation cycle, which again cannot be reconciled with a simple dipolar-like magnetic topology.  
This is also straightfowardly visible from the longitudinal field values that we 
obtain (see Table~\ref{tab:log}), whose temporal variation is inconsistent with 
expectations  from a simple field topology \citep[eg, ][]{wade00b}.  
Although the data at both epochs are too sparse to give a definitive picture of the 
global magnetic topology, we can nevertheless infer that it includes several (presumably radial) 
magnetic field spots of opposite polarities across the equator.  The typical magnetic intensity 
within these spots is a few tens of G (and possibly up to 40~G in the azimuthal field ring), 
significantly greater than the large-scale field strength of the Sun (about 5~G at the pole).  

Further evidence of complexity comes from the apparent variability of the field topology on a time 
scale of a few weeks.  This is directly visible on the reconstructed maps, where the magnetic 
spot distributions at both epochs share little in common.  Although this is not totally unexpected 
given the very incomplete phase coverage, this is also the case around phase 0.50, a portion of the 
stellar surface that was covered at both epochs.  This is particularly obvious in the Zeeman signatures 
collected around phase 0.48 in 2006 June, which are obviously incompatible (both in shape and polarity) with 
that obtained at phase 0.47 in 2006 August and with those obtained at nearby phases (0.38 and 0.55).  
This is further confirmed by trying to fit simultaneously both data sets with a single field topology.  
No acceptable fit to the data can be obtained for values of the period ranging from 11 to 14~d.  
Reduced \chisq\ values lower than 2 can only be obtained for some rotation periods between 12.2 
and 13.4~d. For a period of 12.2~d, for instance, the Zeeman signatures at phases 0.47 and 0.55 in the  
August data set are approximately matched with those at phases 0.30 and 0.38 in the July data set.  
We must therefore conclude that the magnetic topology evolved significantly between our two runs, e.g., 
under the effect of differential rotation. 
If we assume that the variability was not so large as to render the magnetic topologies at both 
epochs fully uncorrelated, we find (by minimising the misfit between the two data sets) that 
the average recurrence period of the Zeeman signatures must be significantly longer than 11.8~d.  

This latter conclusion confirms reports in the literature from photometric monitoring that variations in 
light-curve morphology occur on HD~189733 on timescales of a few tens of days.  As far as our data indicate, 
the recurrence period of the Zeeman signatures is longer than the 11.73-day
period retrieved by 
\citet{croll07}, but shorter than the 13.4-d period found by \citet{winn07b}.  
We suspect that all these discrepant periods actually reflect the fact that HD~189733 is differentially 
rotating;  the recurrence period derived from the various data sets traces essentially the rotation 
period at the latitudes of the main brightness/magnetic features that are present at the surface of the 
star at the time of the observations.  For instance, all data sets could potentially be reconciled by 
assuming that the equator and pole of HD~189733 are rotating with periods of 11.8 and 13.5~d respectively, 
i.e., a latitudinal photospheric shear similar to that of the Sun.  Providing this is indeed the 
correct interpretation, it would imply that differential rotation at the surface of HD~189733 is at 
least as strong as that of the Sun, i.e., with a lap time between the equator and pole of less than 
100~d.  Additional observations are obviously needed to confirm this option;   spectropolarimetric data 
collected over a timescale of about 15~d could settle this issue fairly easily.

\section{Intrinsic stellar vs planet-induced activity}
\label{sec:act}
\begin{figure}
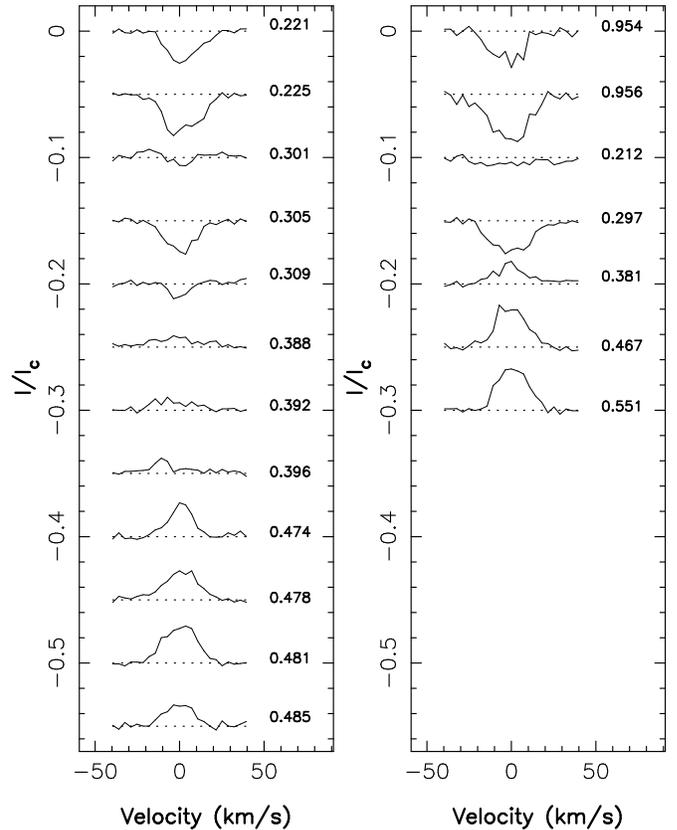

\begin{center}
\epsfig{file=moutou_fig5a.ps,width=11cm,angle=270}
\epsfig{file=moutou_fig5b.ps,width=11cm,angle=270}
\caption{Evolution of \caii\ H\&K residual intensity in the emission core 
(in units of the unpolarised continuum), 
for both the 2006 June (left panel) and August (right panel) data set.  
The corresponding rotation phase is indicated to the right of each profile.  }
\label{fig:cah}
\end{center}
\end{figure}

\begin{figure}
\begin{center}
\epsfig{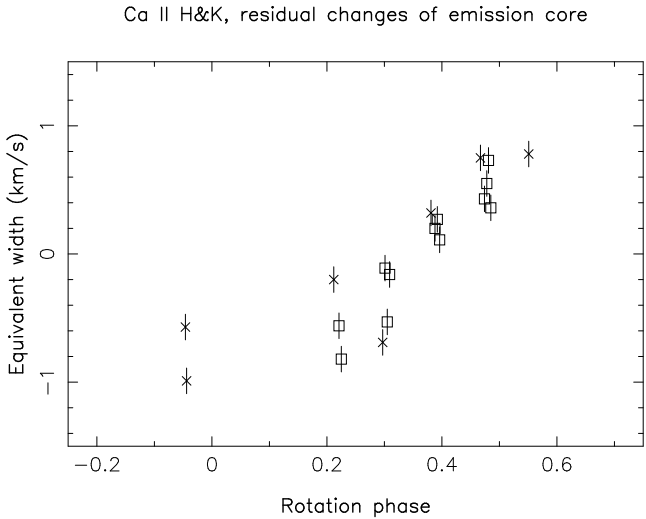}
\caption{Equivalent width variations in the emission cores of the \caii\ H\&K lines 
as a function of rotation phase (derived from the ephemeris of eqn.~\ref{eq:rot}), 
for both the 2006 June (crosses) and August 
(open squares) data sets.  The equivalent width of the average emission core 
is equal to 13~\kms\ in HD~189733.  }
\label{fig:act}
\end{center}
\end{figure}

 HD~189733 is known as a variable star. Its $S_{\rm HK}$ index value is 0.525
  from the catalog of \citet{wright2004}, implying that HD~189733 is among the 10\% 
  most active K dwarfs of the catalog, and among the 20\% most active stars of the 
  whole catalog (including 1,200 stars altogether). 
To investigate how the activity of HD~189733 varies with time, we searched the 
emission cores in the \caii\ H\&K lines for potential variability.  
To maximise the accuracy, we computed a mean profile from both 
\caii\ H\&K lines at each phase, and subtracted from each profile the average 
over the whole data set.  The resulting residuals are shown in Fig.~\ref{fig:cah}.  
Variations of up to $\pm10$\% of the core emission flux are observed throughout both 
runs.  Changes in the amount of core emission are derived by measuring the 
equivalent widths of the residual profiles of Fig.~\ref{fig:cah}.  The 
corresponding values are listed in Table~\ref{tab:log} and plotted in 
Fig.~\ref{fig:act}.  
 When the H and K lines are treated separately, the
  same behaviour is observed in both cores, with similar
  amplitudes. 

The first obvious (and unsurprising) finding is that the activity of HD~189733 
(as witnessed through the \caii\ H\&K emission cores) is intrinsically variable 
on time scales as short as 1~hr. On June 11 for instance, the activity noticeably 
increased between the first and second polarisation sequences, then dropped again 
on the third polarisation sequence.  This sudden burst of activity is probably 
due to a small flare triggered by a local reconnection event;  it is however 
impossible at this stage to speculate about the cause of this event given the 
very sparse data sampling that we have.

\begin{figure}
\begin{center}
\epsfig{file=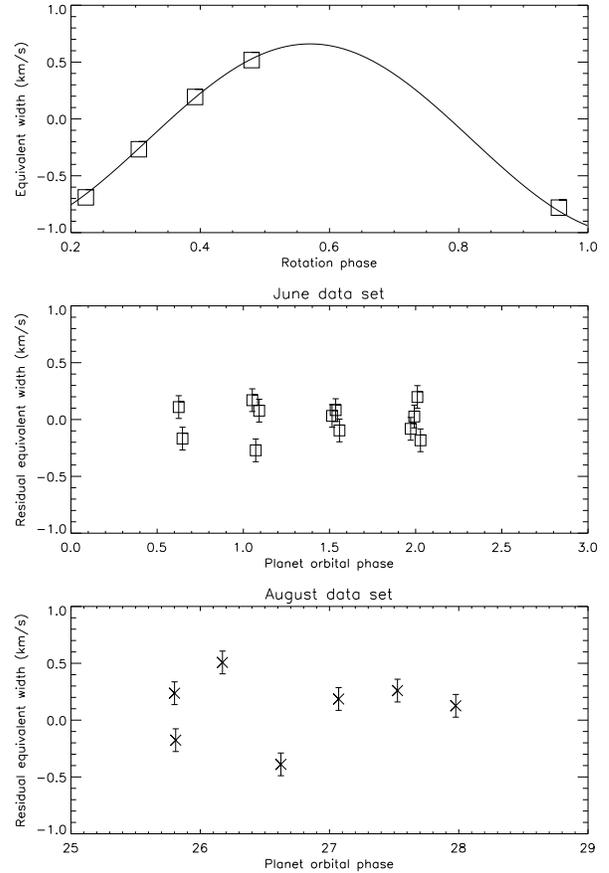,height=12cm}
\caption{Nightly averaged equivalent widths of the \caii\ H\&K lines of the
  June data set as a function of rotation phase fitted by a cosine rotational
  function (top panel); residuals of the individual \caii\ equivalent widths
  after this correction is applied, as a function of the planet orbital phase
  (middle panel for the June data set and top panel for the August data set):
  the high dispersion shows the activity jitter, but at such temporal sampling
  no specific correlation with the planet motion is observed. }
\label{fig:res}
\end{center}
\end{figure}

The longer-term variations in core emission are apparently not related to the orbital phase, 
as one can also see fairly clearly from Table~\ref{tab:log}. While the activity is 
lower than average on June 11 (just after the transit), it is much stronger than 
average about 2 days later, both before, during and after the transit.   
More generally, very little correlation is observed between the activity level and 
the orbital phase, for either data set.  Activity correlates significantly 
better with stellar rotation phase (see Fig.~\ref{fig:act}), though intrinsic 
variability generates a significant dispersion of data points.  
Figure~\ref{fig:res} shows a fit to the nightly-averaged June data (top panel), 
  describing the average rotational modulation in the activity level of HD~189733.  
  Residuals with respect to this average rotational modulation are plotted against 
  the planet orbital phase in Fig.~\ref{fig:res} (lower panels).  
  These residuals both show a large dispersion of individual measurements and no 
  obvious correlation with the planet orbital phase.  The available phase sampling 
  is however still unsufficient;  we thus cannot conclude from our data whether the 
  planet is responsible for activity enhancement along the interaction models 
  proposed in the literature \citep{cranmer,mcivor}.

The variations in \caii\ H\&K emission with stellar rotation phase  appear to be in good mutual agreement for both 
runs(see Fig.~\ref{fig:act}).  Since the activity 
presumably relates to the large-scale magnetic topology, it suggests that the 
intrinsic variability of the field structure as detected through the change in the 
Zeeman signatures (see Sect.~\ref{sec:mod}) is only moderate and local, leaving the 
global field mostly unaffected.  
It also suggests that the rotation period at the footpoints of 
the magnetic loops that confine and heat the chromospheric plasma to produce the observed 
\caii\ H\&K emission is close to 11.8~d, i.e.\ consistent with that derived from 
the MOST photometric data \citep{winn07b} but smaller than the recurrence timescale 
derived from the Zeeman signatures themselves.  This is not necessarily inconsistent;  
Zeeman signatures likely trace both open and closed magnetic field lines, while 
\caii\ emission senses mostly closed field lines.  Assuming that closed field lines 
concentrate around the equator in HD~189733 (as they do in the Sun), this would 
argue for a pole rotating slower than the equator in HD~189733.

Given the incompleteness of our data, these conclusions are of course very uncertain;  
they will however be testable as soon as a data set covering the whole stellar 
rotation cycle is obtained.

Folding the \caii\ H\&K emission variations on half the planet's 
orbital period (ie 1.109~d) produces a 
correlation similar to that of Fig.~\ref{fig:act}.  However, this apparent 
correlation is presumably the fictitious result of aliasing 
between the stellar rotation period and the length of the sidereal day.  Given the sampling of our observations 
(ie roughly one series per day at fixed time, within about $\pm2$~hr), 
modulation on a period of 11.8~d (producing a phase shift of $+8.5\pm0.7$\% 
of the rotation cycle from one day to the next) generates a similar pattern to 
modulation on a period of 1.109~d (producing a phase shift of $-10\pm7$\% 
for two successive observations). The phase diagram by itself is thus not 
sufficient to determine on which timescale activity is modulated.  However, if 
activity was truly modulated on a 1.109~d period, eg due to tidal effects induced by 
the planet, we would then expect other activity indices (e.g. photometry tracing 
surface starspots) to show a similar behaviour;  
according to \citet{winn07b} and \citet{croll07}, no 
modulation on a 1.109~d timescale is observed in their densely-sampled photometric 
data.  We thus conclude that most of the activity of HD~189733 traced by the \caii\ 
emission core is likely varying with the stellar rotation period.  

Note that we do not suggest that activity is completely independent of 
orbital phase in HD~189733;  while the largest fraction of the activity 
level is likely driven by conventional activity processes (and thus modulated 
by the star's rotation), it may well be that a 
small fraction of the observed activity is induced by the planet/star interaction.  
Given the high level of intrinsic variability characterising \caii\ emission 
(see Fig.~\ref{fig:act}), this option can only be checked from observations 
featuring dense temporal sampling over both orbital and rotational periods.  


\section{Transmission spectrum}
\label{sec:atm}

We now consider the transmission signature of the planet's atmosphere in the
intensity spectrum, i.e.,
the residual on-transit spectrum after the off-transit spectrum has been
subtracted. 
One may expect absorption signatures from the
planet atmosphere to be present in the transmission spectrum. Similar attempts
to detect atmospheric spectral features
from the ground have yielded null results on the system HD~209458 
\citep{bundy00,moutou01,brown02,moutou03,snellen04,deming05}, whereas space-based observations gave
positive detections, at a level of 10$^{-4}$ in the visible NaI doublet \citep{charbo02} 
and probably stronger in the UV \citep{AVM03,ballester07}.\\

The on-transit spectrum is the combination of all polarised spectra from sequences 10
and 11, i.e., in total 8 spectra of individual SNR more than 400
at 730 nm. The final combined on-transit spectrum is of SNR
550 to 1100, depending on the wavelength. In the off-transit comparison
spectrum, we included all spectra of the transit night except sequences 10 and 11. 
Each spectrum has to be interpolated to a common spectral format before
combination. The impact of this interpolation should be negligible on features
broader than the spectral sampling interval.
The resulting transmission spectrum does not show absorption features, at a level of 
0.0012 relative flux in the region of the NaI doublet, and more generally, at a level 
between 0.001 and 0.003 over the available spectral range, from 369 nm to 1045 nm.  
Narrow NaI absorption features of 0.45\% amplitude and 0.01nm width, such as predicted 
by models of HD~209458 (Tinetti et al, in prep) are undetected.

\section{Planetary reflected-light spectrum}
\label{sec:ref}

Finally, we searched our linear polarisation spectra for signatures of light from the star
reflected off the planet's atmosphere towards the observer.  
With a short semi-major axis (0.031~AU), a known planetary radius and the geometry of 
the system well constrained by transit observations, the system HD~189733 probably offers 
one of the most favorable configurations to constrain the albedo of a giant exoplanet.  
Upper limits on the albedos of hot Jupiters have been derived from similar searches on HD~75289, 
$\upsilon$~And and $\tau$~Boo, involving several tens of observing nights on 4m to 8m class telescopes 
\citep{charbo99,acc02,leigh03a,leigh03b,leigh03c}.  Their conclusion is that the (assumed grey) albedos 
of hot Jupiters are apparently smaller than 0.12, i.e., much smaller than those of solar-system planets 
like Jupiter and Venus.  
In our case, the star-planet distance is the smallest among all hot Jupiters 
discovered to date orbiting stars bright enough for such observations. It 
thus maximises the strength of the expected reflected signal, making it larger, by at least a factor of 
about 2, than all systems previously observed.  The star is however fainter, compensating at least 
partly for this advantage.  

All previous attempts to detect this spectroscopic signal have been made using unpolarised light only. 
However, this signal is expected to be linearly polarised \citep{seager00, stam04} and can thus also be detected 
through Stokes $Q$ and $U$ spectra.  Detecting this signal through broad-band polarimetry was recently 
attempted with the new PLANETPOL polarimeter \citep{hough06}, with which a potential signal from 
$\tau$~Boo was reported \citep{lucas06} (at a level of about 10~ppm).  Similar measurements can be 
performed through spectropolarimetry.  Since the reflected spectrum is expected to be polarised and 
Doppler shifted as a result of the planet orbital motion, we expect to see tiny polarisation 
structures in the spectrum of HD~189733, oscillating about the main spectral lines of the host star 
and varying in strength with orbital phase.  These polarisation signatures are expected to be maximum 
at quadrature, i.e., for scattering angles of about 90\degr.  

Although spectropolarimetric signatures of exoplanets are potentially easier to identify 
unambiguously (thanks to their modulated Doppler shifts) than broadband polarisation, they are more 
difficult to detect because their amplitudes  are significantly smaller, being reduced by
approximately the relative depth of the average line profile.  
Given the planet's size and orbital distance in the particular case of HD~189733b, we expect the relative broadband 
polarimetric signal to peak at about 50~ppm at most (assuming a grey albedo of 0.4);  we thus expect 
the spectropolarimetric signal to reach about 15~ppm (the relative depth of the LSD profiles being 0.3) 
at orbital phases 0.75 and 0.25.  For a grey albedo of 0.12, this signature is further reduced to a 
maximum size of only 5~ppm.  Our Stokes $Q$ and $U$ spectra each show a relative noise level of 25~ppm;  
we thus need to accumulate between 60 and 600 of them to reach noise levels of 3 to 1~ppm, as is needed to detect the 
spectropolarimetric signal with a precision of at least 5$\sigma$ for albedos ranging from 0.4 down to 0.12 
respectively.  
No useful upper limit on the albedo can therefore be obtained from our single pair of linear polarisation 
spectra;  a few tens of such observations (at the very least) are required to be able to derive a useful 
constraint.

\section{Conclusion}
\label{sec:dis}

HD~189733 is among the most studied planetary systems, with a short-period (2.2d)
Jupiter-like planet transiting the disc of its bright parent star.  At only
0.031~AU distance from its relatively active host star (a K dwarf featuring
moderate rotation), the gaseous planet is embedded in the outer corona and
the magnetosphere of the star.

In order to understand the complex relationships between a star and a
closeby Jupiter-mass planet and in particular to investigate the possible interaction 
that could
lead to planet-induced activity enhancements such as those recently reported in the
literature, we decided to explore the extent and topology of the stellar
magnetosphere of HD~189733.  We used the ESPaDOnS spectropolarimeter at CFHT 
to look for circular polarisation Zeeman signatures in the line profiles of
HD~189733 in 2006 June and August. 

Zeeman signatures are clearly detected in all
spectra, demonstrating that a magnetic field is indeed present at the surface of the 
star.  The Zeeman signatures are not modulated with the planet's orbital period;   
their temporal variation agrees better with the star's rotational modulation,
even though the full rotation cycle of about 11.8~d was not fully covered by our
observations. 

Assuming that the observed variation is mainly caused by rotational modulation,
the detected Zeeman signatures indicate that a large-scale magnetic field of a few tens 
of G is present at the surface of this relatively active K dwarf star.  The reconstructed 
large-scale field topology is significantly more complex than the large scale field of
the Sun, involving in particular a toroidal component and significant contributions from 
magnetic multipoles of order of at least 3.  The magnetic topology is apparently evolving 
on timescales of a few rotation cycles;  moreover, the recurrence timescale of the Zeeman 
signature over the full timescale of our observations is apparently longer than the estimated 
rotation period of 11.8~d.  Both points suggest that significant differential rotation is 
present at the surface of the star.  

The \caii\ H \& K lines clearly feature core emission, whose intensity varies by 
about 10\% on a day-to-day basis.  This variability is mostly modulated with the 
rotation cycle of 11.8~d and presents rapid episodes of intrinsic fluctuations.  The 
overall modulation itself is roughly stable over the timescale of our observations.

At this stage, there is very little we can state concerning putative activity enhancements 
induced by the presence of the giant close-in exoplanet orbiting HD~189733.  In 
particular, its potential contribution to \caii\ H \& K emission needs to be investigated 
through further data, densely sampling both rotation and orbital cycles, to determine  
unambiguously whether, and to which level, the activity of HD~189733 also fluctuates with the planet 
orbital period.  Ultimately, studies such as ours should allow us to diagnose whether 
star-planet interactions rather result from star-planet magnetospheric 
friction (crucially involving the large-scale field of the host star)
or from tidal effects (producing mainly comoving small-scale
magnetic features with no need for a large-scale magnetic field on the host star);  
looking for correlations 
between the planet-induced excess activity and the presence (and topology) of the large-scale field 
should for instance inform us on which of these 2 processes are more likely to occur.

\begin{acknowledgements}

We thank the anonyous referee for constructive comments about an earlier version of this manuscript.  
\end{acknowledgements}

\bibliographystyle{aa}
\bibliography{moutou}

\end{document}